\documentstyle[prd,aps,twocolumn,epsfig]{revtex}
\newcommand{\be}{\begin{eqnarray}}
\newcommand{\ee}{\end{eqnarray}}

\def\lsim{\mathrel{\rlap{\lower4pt\hbox{\hskip1pt$\sim$}}
    \raise1pt\hbox{$<$}}}               
\def\gsim{\mathrel{\rlap{\lower4pt\hbox{\hskip1pt$\sim$}}
    \raise1pt\hbox{$>$}}}               

\begin{document}

\title{
\rightline{\large{Preprint RM3-TH/00-14}}
\vspace{0.5cm}
Comment on "Nucleon elastic form factors and local duality"}

\author{Silvano Simula}
\address{Istituto Nazionale di Fisica Nucleare, Sezione Roma III, Via della Vasca Navale 84, I-00146 Roma, Italy}

\maketitle

\begin{abstract} 

\noindent We comment on  the papers "Nucleon elastic form factors and local duality" [Phys. Rev. {\bf D62}, 073008 (2000)] and "Experimental verification of quark-hadron duality" [Phys. Rev. Lett. {\bf 85}, 1186 (2000)]. Our main comment is that the reconstruction of the proton magnetic form factor, claimed to be obtained from the inelastic scaling curve thanks to parton-hadron local duality, is affected by an artifact.

\noindent PACS numbers: 13.60.Hb;12.38.-t; 13.40.Gp

\end{abstract}

\vspace{0.5cm}

\indent Recently an inclusive electron-proton scattering  experiment \cite{Ioana} has been performed at the Jefferson Lab ($JLab$) in the resonance production region for values of the squared four-momentum transfer $Q^2$ between $\sim 0.45$ and $\sim 3.3 ~ (GeV/c)^2$. The aim was to investigate the connection among the resonance and the scaling regions, known as parton-hadron local duality \cite{BG}. The new data were found to exhibit the local duality for each of the most prominent proton resonances. In \cite{Ioana} a fit to the average strength of all the resonances was carried out and, thanks to the parton-hadron local duality, interpreted as the {\em scaling} curve. Here below, we refer to such a fit as the $JLab$ fit, viz.
 \be
      F_2^{(JLab)}(\xi) & = & \xi^{0.870} (1 - \xi)^{0.006} \cdot \left[ 
      0.005 - \right. \nonumber \\
      & &  \left. 0.058 (1 - \xi) - 0.017 (1 - \xi)^2 + \right. 
      \nonumber \\
      & & \left. 2.469 (1 - \xi)^3 - 0.240 (1 - \xi)^4 \right]
      \label{eq:JLABold}
 \ee
where $\xi \equiv 2 x / [1 + \sqrt{1 + 4 M^2 x^2 / Q^2}]$ is the Nachtmann variable, which includes the effects of target-mass corrections, improving at finite $Q^2$ the Bjorken scaling variable $x$. In order to constrain the large-$\xi$ behavior of the $JLab$ fit the authors of \cite{Ioana} have employed $SLAC$ data up to $Q^2 = 8 ~ (GeV/c)^2$. This means that the highest $\xi$-point constraining the $JLab$ fit is $\xi \simeq 0.86$, corresponding to the $\Delta(1232)$ location at $Q^2 = 8 ~ (GeV/c)^2$.

\indent An interesting question is whether local duality may be applied to the proton elastic peak \cite{BG,SIM,Georgi,Ricco}. If local duality holds also in the unphysical region extending up to $\xi = 1$ (which corresponds at finite $Q^2$ to $x > 1$), the proton magnetic form factor $G_M^p(Q^2)$ can be obtained from the moment of order $n$ of the scaling function, $F_2^p(\xi)$, viz. (cf. \cite{SIM})
 \be
        G_M^p(Q^2) =  \sqrt{ {2 - \xi_{el} \over \xi_{el}^n}  \mu_p^2 {1 
        + \tau \over 1 + \mu_p^2 \tau} \int_{\xi_{\pi}}^{\xi^*} d\xi 
        \xi^{n-2}  F_2^p(\xi) }
       \label{eq:GMp}
 \ee
where $\mu_p$ is the proton magnetic moment, $\tau =$ $Q^2 /$ $4 M^2$, $\xi_{el} =$ $2 / [1 + \sqrt{1 + 1 / \tau}]$, $\xi^* =$ $min[1, Q/M]$ and $\xi_{\pi}$ is the pion production threshold. Note that $\xi_{\pi} (\xi_{el}) =$ $0.41 (0.50),$ $0.63 (0.70),$ $0.76 (0.81),$ $0.83 (0.87),$ $0.90 (0.92)$ at $Q^2 =$ $0.45,$ $1.4,$ $3.0,$ $5.0,$ $10$ $(GeV/c)^2$, respectively. In \cite{Ent}, adopting for $F_2^p(\xi)$ the $JLab$ fit (\ref{eq:JLABold}) and considering only $n = 2$, the reconstructed $G_M^p(Q^2)$ was shown to agree with the data within $30 \%$ up to $Q^2 \sim 7 ~ (GeV/c)^2$. This result is at variance with the findings of Refs. \cite{BG,SIM,Georgi,Ricco}.

\indent We start noting that in the righthand side of Eq. (\ref{eq:JLABold}) the term proportional to $(1 - \xi)$, which any way is not consistent with quark counting rules, has a negative coefficient, so that $F_2^{(JLab)}(\xi)$ may be a monotonic increasing function at large $\xi$. This is indeed the case as shown in Fig. 1 by the dashed line\footnote{Note that in Eq. (\ref{eq:JLABold}) the term $(1 - \xi)^{0.006}$ ensures that $F_2^{(JLab)}(\xi = 1) = 0$, but in practice it has no effect at all for $\xi$ up to $0.9999$, as it can be easily checked numerically.}, which exhibits an anomalous behavior at $\xi \gsim 0.9$, i.e. beyond the highest $\xi$-point constraining the $JLab$ fit of \cite{Ioana}.

\begin{figure}[htb]

\centerline{\epsfxsize=7.50cm \epsfig{file=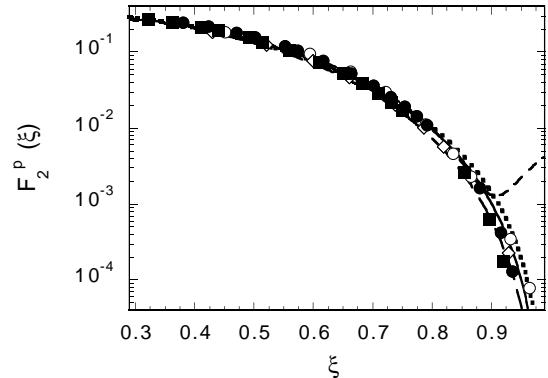}}

\caption{The proton scaling function $F_2^p(\xi)$ versus the Nachtmann variable $\xi$. The dashed and solid lines correspond to the $JLab$ fit (\protect\ref{eq:JLABold}) of \protect\cite{Ioana} and to our modified $JLab$ fit given by Eq. (\ref{eq:JLABnew}), respectively. The open dots and diamonds, and the full dots and squares are the average strengths, obtained using the $SLAC$ parameterization \protect\cite{SLAC} of the proton structure function, in the $\Delta(1232)$, $S_{11}(1535)$, $F_{15}(1680)$ and "higher-mass" resonance regions, as defined in \protect\cite{Ioana}, respectively. The long-dashed and dotted lines correspond respectively to the $GRV$ set \protect\cite{GRV} of $PDF$'s and to the $NMC$ parameterization \protect\cite{NMC}, omitting for the latter its $1 / Q^2$ term (see text), evaluated at $Q^2 = 10 ~ (GeV/c)^2$.}

\end{figure}

\indent In order to clarify the impact of the anomalous shape of the $JLab$ fit on the reconstruction of $G_M^p(Q^2)$ through Eq. (\ref{eq:GMp}), we have simply developed  a {\em modified} version of the $JLab$ fit, which coincides with the original one within $\pm 10 \%$ for $\xi \lsim 0.86$ , but exhibits a monotonic decreasing behavior at larger $\xi$, viz.
 \be
      \tilde{F}_2^{(JLab)}(\xi) & = & \xi^{0.940} \left[ 2.650 (1 - 
      \xi)^{3.38} + \right. \nonumber \\
      & & \left. 0.240 (1 - \xi)^4 \right]
     \label{eq:JLABnew}
 \ee

\indent In Fig. 1 the modified $JLab$ fit is reported as the solid line and compared with the average strengths of the most prominent proton resonances, generated using the parameterization of the inelastic $SLAC$ data of \cite{SLAC}. It can be seen that our modified $JLab$ fit is in reasonable agreement with the $SLAC$ resonance averages up to very large values of $\xi$. Finally, the proton structure function $F_2^p(\xi)$ evaluated at $Q^2 = 10 ~ (GeV/c)^2$ using the $GRV$ set \cite{GRV} of parton distribution functions ($PDF$'s) and the $NMC$ parameterization of \cite{NMC} is shown in Fig. 1. Note that the $NMC$ fit contains an explicit power correction term proportional to $1 / Q^2$, which has been excluded. Indeed, as shown in \cite{SIM}, the replacement of the Bjorken variable $x$ with the Nachtmann variable $\xi$ is an approximate way to consider target mass ($TM$) corrections at large $Q^2$. Therefore, the inclusion of the $1 / Q^2$ term of the $NMC$ fit (which incorporates already $TM$ effects) and, at the same time, the use of the variable $\xi$ would lead to an overcounting of $TM$ effects. From Fig. 1 it can be seen that up to $\xi \simeq 0.85$ the shape of the $JLAB$ fit is not inconsistent with standard $PDF$ expectations as well as with the $NMC$ fit, provided $TM$ effects are properly included. This is at variance with the results reported in Fig. 2 of \cite{Ioana}. There, however, $TM$ effects were not included consistently in the curves labeled "MRS(G)" and "CTEQ4". We have checked that, after proper inclusion of $TM$ corrections, those curves become close to the results labeled "NMC" (cf. also Fig. 3 of \cite{SIM}).

\begin{figure}[htb]

\centerline{\epsfxsize=7.50cm \epsfig{file=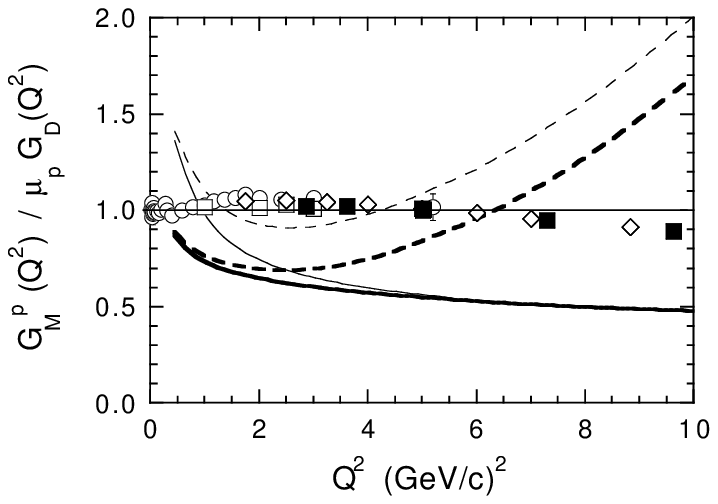}}

\caption{The proton magnetic form factor $G_M^p(Q^2)$, divided by its dipole expectation $\mu_p G_D(Q^2) \equiv 2.793 /$ $(1 + Q^2 / 0.71)^2$, versus $Q^2$.  Open dots, squares, diamonds and full squares are the experimental data from Ref. \protect\cite{data}(a), (b), (c) and (d), respectively. The dashed and solid lines are the results of Eq. (\protect\ref{eq:GMp}) obtained using the $JLab$ fit (\protect\ref{eq:JLABold}) and its modified version (\protect\ref{eq:JLABnew}), respectively. Thick and thin lines correspond in Eq. (\protect\ref{eq:GMp}) to $n = 2$ and $n = 10$, respectively.}

\end{figure}

\indent In Fig. 2 the proton magnetic form factor $G_M^p(Q^2)$, resulting from the application of the parton-hadron local duality [Eq. (\ref{eq:GMp})] using the original and our modified $JLab$ fits, is shown and compared with the data. In evaluating Eq. (\ref{eq:GMp}) we have considered both $n = 2$ and $n = 10$. The former case is the only one employed in \cite{Ent}, while the latter is representative of the case of higher moments which are more sensitive to the shape of the scaling curve at large $\xi$. It can be seen that : ~ i) the results we have obtained using the $JLab$ fit (\ref{eq:JLABold}) coincide with those of \cite{Ent} for $n = 2$ (thick dashed line), but exhibit a remarkable dependence on the order $n$ of the moment (compare thin and thick dashed lines); ~ ii) the anomalous shape of the $JLab$ fit (\ref{eq:JLABold}) heavily affects the reconstruction of $G_M^p(Q^2)$ for $Q^2 \gsim 5 ~ (GeV/c)^2$ (compare thick dashed and solid lines); ~ iii) using our modified $JLab$ fit the values of $G_M^p(Q^2)$ obtained via the application of parton-hadron local duality, underestimates the data by a factor of $\simeq 2$ for $Q^2 \gsim 2 ~ (GeV/c)^2$ (see thick solid line); ~ iv) for $Q^2 \lsim 2 ~ (GeV/c)^2$ the reconstructed $G_M^p(Q^2)$ appears to be close to the experimental data only if $n = 2$ is adopted (compare thin and thick lines).

\indent To sum up, the main conclusion of Ref. \cite{Ent}, concerning the possibility of reconstructing the proton magnetic form factor from the inelastic scaling curve, is the result of an artifact in the $JLab$ fit (\ref{eq:JLABold}) of Ref. \cite{Ioana}. Using our modified $JLab$ fit [see Eq. (\ref{eq:JLABnew})] we have shown that the application of the parton-hadron local duality, as given by Eq. (\ref{eq:GMp}), fails to reproduce existing data on $G_M^p(Q^2)$ at least for $Q^2$ up to $\sim 10 ~ (GeV/c)^2$, in agreement with the findings of Refs. \cite{BG,SIM,Georgi,Ricco}.

\end{document}